\documentstyle[sprocl]{article}

\def\be{\begin{equation}}
\def\ee{\end{equation}}
\def\bea{\begin{eqnarray}}
\def\eea{\end{eqnarray}}

\begin{document}

\title{PUZZLING ASPECTS OF HOT QUANTUM FIELDS}

\author{T. GRANDOU}

\address{Institut Non Lin\'eaire de Nice, 1361 route des lucioles,\\
06560 Valbonne, France \\
E-mail: grandou@inln.cnrs.fr}

\maketitle\abstracts{ Over the past decade, Finite Temperature  
Quantum Field Theories (FTQFT's) have benefitted
from impressive developments, while an increasing number of  
intriguing points were made. Some of them are
presented
here, recent and older, in a non exhaustive list.}

\section{In the beginning..}

Soon after the Asymptotic Freedom property of the strong running  
coupling constant of QCD was established,
Collins
and Perry
  \cite {colinpery} suggested that the same property holds true in  
ultra dense nuclear matter conditions as well.
  The idea was then quickly extended to conditions where the  
effects of not so high a matter density $\mu$, could
  be
  compensated for by a high enough temperature $T$, the whole  
picture boiling down to a diagram displaying the
confined and
   deconfined phases of the QCD quanta. In the same line of  
reasoning, it is to be noted that, likewise, broken
   symmetries are expected to be restored in the high $T$-limit  
\cite {lindolaja}. However, just the opposite
   scenario was also shown to be a possible outcome \cite  
{weinbergpisarski}!

\section{An axiomatic point of view}

One can show that FTQFT's can be renormalized {\it{\`a la}} BPHZ,  
and that, once equations of motion,
standard axiomatic requirements (such as space time translation and  
rotational invariances, locality, cluster
property, etc..), and the KMS conditions specific to thermal  
equlibrium are imposed, then, the perturbative
expansion of the (Grand Canonical) thermal correlation functions
$$W(x_1,x_2, .. ,x_n)=<\varphi(x_1)\varphi(x_2) .. \varphi(x_n)>_\beta$$
enjoy a
unique determination, and allow a reconstruction of the full  
representation space of the field algebra by means
of
 the GNS construction \cite {steinmann}. On the other hand, these  
thermal correlation functions may as well not
 exist at all in view of Infra-Red (IR) non (Lebesgue)-integrable  
singularities related to vanishing external
 momenta, and/or exceptional combinations of them! Such IR non  
integrable singularities
 have effectively been met in the course of actual perturbative  
calculations \cite {grandlebel}. A
 natural question is thus: could these IR non integrable  
singularities be compensated by other diagrams, as
 thermal Gauge Field Theories have proven to be pretty rife with  
cancellations? In the case of two point
functions, for  example, it has been shown recently that for a  
large enough variety of self energies and
vertices, self energy insertions along internal lines cancel
 against insertions of {\it{rungs}} in ladder type diagrams \cite  
{leb.smilg.carr.kob.petit}. However this nice
and
 simplifying mechanism is effective in some kinematical regimes  
only where a particular resolution of a
 Schwinger-Dyson equation for the vertex, in terms of the self  
energy function, is available. What  about  other
phase  space  regimes?  And what about the calculation of processes  
related to thermal
correlation functions with a higher number of
external
 momenta, some of them exceptional?

\section{Perturbative Regimes}
Analyzing the representation space of the GNS construction, it has  
been shown that (due to Lorentz
invariance explicit thermal breaking) FTQFT's are inherently non  
perturbative in the following sense: there are
no $LSZ$-asymptotic states in term
of which to devise any kind of $S$-matrix approach \cite  
{landsman}. Only quasiparticle states could be used for
that purpose, but unfortunately, such attempts quickly become  
practically untractable. This,
however, is just one particular realization of  
"{\it{Non-Perturbativeness}}" a notion which, being negative,
opens up over other
aspects and realizations, whose inter-relations are not necessarily  
well understood or even investigated. A more
immediate aspect is of course related to the smallness of the  
effective running coupling constant. In this
respect,
a large number of calculations of $\alpha_s(\mu, T)$, the  
dependance on the temperature $T$ of the effective
running strong coupling of QCD, have been performed, with rather  
odd results. Different from each others, the
results concerning the GellMann-Low $\beta_T$-function, can all be  
written as
\begin{equation}\beta_T(\alpha_s(\mu, T))= T\frac  
{\partial}{\partial T}(\frac {g^2_R}{4\pi}=\alpha_s(\mu,
T))=-C(T)\ \alpha^2 _s(\mu, T)\label{eq:betaT}\end{equation} where  
$\mu$ stands for some $T=0$ renormalization
scale. The prefactors $C(T)$ came out dependent, (i): on the vertex  
choosen to define the renormalized coupling
(tri-or quadri-linear in the gauge fields, gauge-quarks,  
ghost-gauge vertex), (ii): on the external momenta
vertex configurations, (iii): .. and on the gauge choice itself! No  
meaning could accordingly be attached to
these
results. The most satisfying effort produced in order to get rid of  
the above drawbacks is certainly the
one
having made use of the so called Vilkovisky-DeWitt effective  
action, $\Gamma_{VD}$, which, contrarily to the
conventional effective action of QCD \cite{rebhan}, is explicitly gauge
invariant and independent of the condition choosen to quantize the  
theory. An unambiguous procedure results where
the renormalized coupling normalization conditions are entirely  
fixed by gauge invariance itself, and any
drawback (i) to (iii) is consistently circumvented. Now, the result  
can essentially be written as
\begin{equation}
\beta_T(\alpha_s)=-\left(b-\frac{21}{6}\pi^2N_C(\frac{T}{\mu})+{\cal{O}}((\frac{T}{\mu})^{-2})\right)
\alpha^2_s\label{eq:betaTVD}\end{equation}
where $b=(11N_C-2N_f)/6$. For large enough $T's$, the  
$\beta_T$-function is positive, and, in contrast with
Lattice data as well as physical intuition,
Asymptotic Freedom at high $T$ (and/or density) cannot be deduced  
from one loop Renormalization Group arguments!
It
seems hard to believe that the one loop specificity could be at  
fault in this somewhat disappointing result, and
more satisfying to keep in mind that $\Gamma_{VD}$ is not, itself,  
free of any drawback. For example, its
expansion
in terms of mean fields $\overline{A}$ (and $\overline{A}_0$, in  
the absence of sources) is
\begin{equation}
\Gamma_{VD}(\overline{A})= \Sigma
\frac{1}{n!}{\Gamma_{VD}^{(n)}}^{i_1..i_n}(\overline{A}_{i_1}-\overline{A}_{0i_1}
)\ ..\ (\overline{A}_{i_n}-\overline{A}_{0i_n})  
\label{eq:volterra}\end{equation}and the coefficients,
$\Gamma_{VD}^{(n)}$ cannot be interpreted in terms of
1PI-correlation function of $A$ or of any other operator. This is  
not the case in a Wilsonian flow approach of
hot QCD, where gauge invariance can also be maintained along the flow
\begin{equation}\delta_\alpha \  
\{\Gamma_{k,T}(\overline{A})-\Gamma_{k,T=0}(\overline{A})\}= 0  
\label{eq:wilsonian}\end{equation}where
the cutoff $k$ is the flow parameter \cite {litim}. Certainly, it  
would  be interesting to see the results one
would
get for $\beta_T$.

\section{The Resummation Program}

From now on, we assume a small enough $\alpha_s$. For Green's  
functions with soft, order $gT$-external momenta,
a
re-organization of bare Perturbation Theory (PT) is mandatory to  
get the completeness of leading thermal
corrections. This is achieved through a Resummation Program (RP)  
\cite{bratenpisarski}, a beautiful effective
Perturbation
Theory,
ruling soft scale fluctuations, and a leading order approximation  
scheme, fully consistent with gauge
invariance. In particular, the RP has solved the static gluon  
damping rate problem, and thanks to
the
{\it{Dynamical Screening Mechanism}} \cite{lebellac}, seems to have  
improved the IR sector of bare PT. It is
however
from damping rate's calculations that two major obstructions to the  
RP came about, with first \cite{pis88}, the
moving fermion
damping rate of QED and QCD. In effect, when the latter is  
evaluated by the fermion mass shell, a logarithmic IR
singularity is found in both QED and QCD cases
\begin{equation}
\gamma(E,p)_{|_{E=p}}=\frac {e^2T}{2\pi}\int_{\frac  
{|E-p|}{2}}^{k^\star} \frac {kdk}{k^2}\ +\ regular
\label{eq:gamma}\end{equation}where the fermion has been taken  
massless for the sake of simplicity, while the
same
holds true for massive fermions too. This IR singularity is due to  
unscreened transverse modes, as can be read
off the intermediate energy sum rule leading to ({\ref{eq:gamma}})
\begin{equation}\int_{-k}^{+k} \frac{dk_0}{2\pi k_0}\  
{}^\star\rho_T(k_0,k) \simeq
\frac{1}{k^2}+{\cal{O}}(\frac{1}{m_D^2}),\
\ \ \ \ \ k/m_D
<<1,\ \ \  
m^2_D={\cal{O}}(g^2T^2)\label{eq:sumrule}\end{equation}where  
${}^\star\rho_T$ is the transverse
spectral
density associated to the RP's
effective propagator (\ref{eq:efprop}). A further (Bloch-Nordsieck  
type) resummation has been proposed in order
to screen this
IR singular behaviour \cite{blaiziancu}. A possible transverse  
screening mass has also been looked for in QCD,
with
some encouraging result of order $g^2T$, in the case of a three  
dimensional $SU(N_C)$ theory \cite {nair}.
However
it has been recently argued that such a mass could be too small to  
act as an efficient IR cutoff in the case of
QCD \cite
{gelis}, and eventually, no such mass can be invoked for QED. One  
may therefore wonder how the problem could get
translated in another resummation scheme of the thermal leading  
effects, as we will comment shortly.
The emission rate of a soft real photon from a quark-gluon plasma  
has also been found troublesome \cite {bps},
since, up to regular terms, one gets a result affected with a  
collinear singularity (the dimensional
$\varepsilon$
parameter regularizes the divergence)  \begin{equation}  
\frac{C^{st}}{\varepsilon}\int
\frac{d^DP}{(2\pi)^D}\delta({\widehat{Q}}\cdot  
P)(1-2n_F(p_0))\sum_{s=\pm 1,R=P,P+Q}\pi(1-s\frac{r_0}{r})\beta_s(R)
\label{eq:pising}\end{equation}
where $\beta_s(R)$ is related to the fermionic effective  
propagator. A so-called {\it{Thermal Asymptotic Mass}},
$m_\infty$, has been proposed a gauge invariant way  
\cite{flechsigrebhan}, of order $gT$, which takes the original  
$1/\varepsilon$ of (\ref{eq:pising}) to a large
logarithm $\ln(1/g)$. However the method, if consistent, does not  
really save the general situation as we will
see shortly, so that again, one may wonder about what could come  
out of another resummation scheme of the
thermal leading effects. We come to this point now.

\section {Perturbative Resummation}
A Perturbative Resummation Scheme of the thermal leading effects  
can be introduced \cite {candelp},
differing the
usual RP, only the effective propagators. Instead of effective  
functions like
\begin{equation}{}^\star\Delta_{\alpha\alpha}
(K)=\frac{i}{K^2-\Pi^{HTL}_{\alpha\alpha}(k_0,k)+i\varepsilon_\alpha}
\label{eq:efprop}\end{equation}with $\alpha$ labelling the two  
Retarded/Advanced
field types of a (R/A) real time formalism for example, and with  
the superscript $HTL$ to mean the leading part
of
$\Pi$, order $g^2T^2$, we use the geometrical series representations
\begin{equation}\sum_{N=0}^\infty
\Delta^{(N)}_{\alpha\alpha}(k_0,k)=i\sum_{N=0}^\infty
(\Delta^{(0)}_{\alpha\alpha}(k_0,k))^{N+1}
\left(\Pi_{\alpha\alpha}^{HTL}(k_0,k)\right)^N
\label{eq:series}\end{equation}
effective vertices remaining the same in both RP and Perturbative  
Resummation scheme. The effective
functions (\ref{eq:efprop}) and (\ref{eq:series}) satisfy the same  
Dyson equation,
\begin{equation} {}^\star\Delta_{\alpha\alpha}
(k_0,k)=\Delta_{\alpha\alpha}^{(0)}(k_0,k) +  
\Delta_{\alpha\alpha}^{(0)}(k_0,k)\Pi_{\alpha\alpha}^{HTL}(k_0,k){}^\star\Delta_{\alpha\alpha}
(k_0,k)\label{eq:dyson}\end{equation}
 but are definitely different one
dimensional $k_0$-distributions. When used in the  course of  
practical calculations, they lead to
irreducibly different IR
behaviours, and this feature appears specific to the thermal  
context \cite {candelp}. While (\ref{eq:series}) is by
construction a Taylor series in $g^2$, (\ref{eq:efprop})
is not; this is why the sum rule (\ref{eq:sumrule}) for example,  
can be seen to display a Laurent series
behaviour in
$g^2$. In the moving fermion damping rate calculation, the outcome  
is that the series associated to longitudinal
degrees is analytic in a domain whose real restriction is given by  
$|\vec{k}|\geq m_D$, whereas transverse
degrees
of freedom rigorously do not contribute
\begin{equation}\sum_{N=0}^\infty \int_{-k}^{+k}\frac{dk_0}{\pi
k_0}(1-\frac{k_0^2}{k^2})\ {\rm{disc}}_{p_0}\  
\Delta_{\alpha\alpha}^{(N)}(k_0,k)=0
\label{eq:nomagnet}\end{equation}There are no unscreened  
magnetostatic modes problem in this resummation scheme,
whereas the same IR singularity as displayed in (\ref{eq:gamma})  
results of a continuation to $0\leq k\leq m_D$
of
the longitudinal series; and the same features here equally apply  
to both QED and QCD. For the soft real photon
emission rate, or second obstruction (\ref{eq:pising}), a very  
different setting of the problem is obtained also
within the Perturbative Resummation
scheme \cite{thierry}.

\section {Completeness}

Though collinear divergences are effectively regularized by the  
introduction of a {\it{Thermal Asymptotic Mass}},
the latter, we wrote, "does not save". This is because of the  
{\it{Collinear Enhancement Mechanism}} discovered a
few years ago \cite{petitgirard}. This mechanism shows up in the  
calculation of processes related to Green's
functions with
external momenta on the light cone, and makes higher number of  
loops contributions as large, or even larger than
lower ones, thus compromising the RP's completeness. One has  
symbolically, with $\hat{K}$ denoting the light-like
vector $(1,\hat {k})$
\begin{equation}
\int  
\frac{d\hat{K}}{4\pi}\frac{{\cal{F}}(\hat{K},P,R,..)}{(\hat{K}\cdot  
P -\frac{m^2_\infty}{2k})(\hat{K}\cdot R
-\frac{m^2_\infty}{2k})\ ..\ (\ .. \ )}\ \ \longrightarrow\ \   
(\frac{1}{g})^n F(P,R,..)\label{eq:amplification}
\end{equation}
This in the end, should not come out as a
too big surprise if
we keep in mind that, first devised in the imaginary time  
formalism, the RP's power
counting analysis implicitly assumed
angular integrals on the order of unity! Now, if we are more or  
less used to the idea that the RP breaks down by
the light cone, we may be not so familiar with the idea that it  
could be so elsewhere too. However, if we take
the
static photon
production rate out of a Quark-Gluon Plasma, we find the one loop  
result \cite{bratenpisyuan}
\begin{equation}  
\rm{Im}\Pi_R(q_0,\vec{0})=-\frac{e^2g^2N_cC_F}{32\pi} q_0T
\ln\left(\frac{2q_0T}{Max\{q_0^2,m_F^2\}}\right)\label{eq:yuan}\end{equation}
where $m_F^2=g^2T^2C_F/8$ is the fermionic thermal mass. Now, as  
shown more recently \cite{augelpekozar}, a two
loop contribution is
\begin{equation}  
\rm{Im}\Pi_R(q_0,\vec{0})=-\frac{e^2g^2N_cC_F}{32\pi} q_0T
\ln\left(\frac{2q_0T}{q_0^2+m_g^2}\right)\label{eq:augelpekozar}\end{equation}where  
$m_g$ is the gluonic Debye
mass
of order $gT$. The striking aspect of (\ref{eq:yuan}) and  
(\ref{eq:augelpekozar}) is of course that the same
orders
of magnitude and dependences on $q_0$ and $T$ are obtained at one  
and two loops, and similar contributions will
also have to be looked for at any arbitrary higher
number of loops. This new RP's breakdown takes place out of the  
light cone region.

\section {Low energy limit}

To end up with these Hot Quantum Fields somewhat puzzling aspects,  
let us mention that the low energy limit of
Hot Gauge Theories (that is, at an energy scale on the order of  
$g^2T$ at most), expected to be both local
and
stochastic, has effectively been found so in a series of recent  
papers \cite{arbodeiasonyaf}. This however holds
true
at a leading-logarithm level
of approximation, while lattice calculations would rather indicate  
very large subleading-logarithm
corrections \cite{humomu}!

\section {Conclusion}

Compared to their ordinary vacuum representations, the quantized  
field algebras-$KMS$ (thermal) representations
display an unexpected richness of new structures and results, all  
of them being eventually generated by the interplay of
only two (non decorrelated) major features: (i) The Lorentz  
invariance thermal breaking, and, (ii) The appearance of a
dimensionful parameter in the formal perturbative series, namely the
temperature. The few puzzling points which we have listed here,  
display the surprising richness of the subject as
well as its rather unexpected difficulties. It may be, as some  
authors are inclined to think, that we are running
short of some deeper conceptual understanding of FTQFT's. In any  
case, we are certainly lacking clear cut
experimental results that would help us knowing wether we are going  
the right way or not .. including the very
first of all steps, implementing the thermodynamical temperature on  
the basis of a formal analogy with an imaginary
time \cite{bloch}
\begin{equation}
it/ \hbar \longleftrightarrow 1/k_BT \label{eq:analogy}\end{equation}
though it is not impossible either, that this formal analogy itself  
relies on a much deeper and fundamental duality
relation between time and absolute temperature \cite{krol}, .. a  
possibility which would certainly stand for one
more intriguing aspect of FTQFT's!

\section*{Acknowledgments} I wish to express my gratitude to H.M. Fried
and B. M$\ddot{\rm{u}}$ller for having organized this fifth QCD
workshop.

\end{document}